\documentclass[preprint,notoc]{JHEP3}
\usepackage{epsfig}

\def\to{\rightarrow}

\def\bi{\begin{itemize}}
\def\ei{\end{itemize}}

\def\tst{\tilde t}
\def\ttau{\tilde \tau}

\def\tell{\tilde\ell}
\def\tq{\tilde q}

\def\tw{\widetilde W}
\def\tz{\widetilde Z}
\def\alt{\stackrel{<}{\sim}}
\def\agt{\stackrel{>}{\sim}}
\def\be{\begin{equation}}  
\def\ee{\end{equation}}

\title{Target dark matter detection rates \\
in models with a well-tempered neutralino}
\author{Howard Baer$^{a,b}$, Azar Mustafayev$^c$, Eun-Kyung Park$^b$ 
and Xerxes Tata$^d$\\
$^a$Department of Physics, University of Wisconsin, Madison, WI 53706, USA\\
$^b$Department of Physics, Florida State University, Tallahassee, 
FL 32306, USA\\
$^c$Department of Physics and Astronomy, University of Kansas,
Lawrence, KS 66045, USA\\
$^d$Department of Physics and Astronomy, University of Hawaii,
Honolulu, HI 96822, USA\\
E-mail: \email{baer@hep.fsu.edu}, \email{amustaf@ku.edu},
\email{epark@hep.fsu.edu}, \email{tata@phys.hawaii.edu}}

\preprint{\vbox{FSU-HEP-061018, UH-511-1097-06}}

\abstract{In the post-LEP2 era, and in light of recent 
measurements of the cosmic
abundance of cold dark matter (CDM) in the universe from WMAP, many
supersymmetric models tend to predict 
1. an overabundance of CDM and 
2. pessimistically low rates for direct detection of neutralino dark matter.
However, in models with a ``well-tempered neutralino'', where the 
neutralino composition is adjusted to give the measured abundance of CDM, 
the neutralino is typically of the mixed bino-wino
or mixed bino-higgsino state.
Along with the necessary enhancement to neutralino annihilation rates, 
these models tend to give elevated direct detection scattering rates compared 
to predictions from SUSY models with universal soft breaking terms.
We present neutralino direct detection cross sections from a variety of
models containing a well-tempered neutralino, and find 
cross section asymptotes
with detectable scattering rates. 
These asymptotic rates provide targets that various
direct CDM detection experiments should aim for.
In contrast, in models where the neutralino mass rather than its
composition is varied to give the WMAP relic density
via either resonance annihilation or co-annihilation, 
the neutralino remains essentially bino-like, and  
direct detection rates may be below the projected reaches of all proposed
experiments.}


\begin{document}

\section{Introduction}

One of the compelling successes of $R$-parity conserving 
supersymmetric models is the prediction
of a candidate particle to account for cold dark matter (CDM) in the universe.
The lightest neutralino $\tz_1$ is especially attractive\cite{haim,griest}, 
since it could be produced 
thermally in the early universe with a cosmic abundance of the right order
of magnitude to match precise measurements by the WMAP collaboration
combined with data from the Sloan Digital Sky Survey which
yield \cite{wmap}
\be
\Omega_{\rm CDM}h^2 = 0.111^{+0.011}_{-0.015} \ \ (2\sigma)\ .
\label{eq:Oh2}
\ee
The CDM relic abundance can be predicted in particle physics models with
thermal WIMPs (such as the stable neutralino of supersymmetric models),
where it is found that, aside from the additional complication of
possible co-annihilation with electrically charged or colored sparticles
or accidental resonance enhancements,
\be
\Omega_{\rm WIMP}h^2\sim \frac{0.1\ {\rm pb}}{\langle\sigma v_{rel}\rangle}
\sim 0.1 \left(M_{\rm SUSY}\over{100 \ {\rm GeV}}\right)^2\;,
\label{eq:susypred}
\ee
where $\langle\sigma v_{rel}\rangle$ is the thermally averaged
neutralino annihilation cross section times relative velocity, and
$M_{\rm SUSY}$ is the sparticle mass scale. Assuming no hierarchy in the
sparticle spectrum, we see that sparticles with weak scale masses give
the correct order of magnitude~(\ref{eq:Oh2}) for the relic density,
whereas for much larger sparticle masses the predicted relic density
will be too large {\it unless} the neutralino annihilation cross section
in the early universe is enhanced from its naive value.
The smallness of the error bars on the CDM relic density measurement
provides a stringent upper bound on the relic CDM abundance predicted by
supersymmetric models.\footnote{Throughout our analysis, we assume
thermal production of neutralinos and standard Big Bang cosmology, even
at very early times in the history of the Universe. We recognise that it
is possible to build phenomenologically viable models where the very early
history of the Universe is significantly altered. In these more
complicated cosmologies, 
our considerations would not apply \cite{graciela}.} The
lower bound is less certain, since the dark matter may be comprised of
several particles
and the neutralino need not saturate the value
in (\ref{eq:Oh2}).

In early analyses of supersymmetric dark matter,
the favored neutralino annihilation mechanism in the
early universe was taken to be $\tz_1\tz_1 \to f\bar{f}$ (where $f$ is a 
SM fermion), which occurs via $t$-channel sfermion exchange. 
Many analyses were performed within constrained frameworks 
where squark, slepton and gaugino mass parameters are related at 
some high energy scale, and where the sleptons tend to be lighter than
squarks owing to renormalization group effects.
Within such models, neutralino annihilation to leptons then has a larger cross
section than annihilation to quarks since $m_{\tell} < m_{\tq}$.
Assuming sfermion exchange as the dominant neutralino annihilation mechanism,
the rather low value of $\Omega_{CDM}h^2$ measured by WMAP
favors quite light sparticle masses $\sim 100$~GeV. 
At the same time, sparticle search limits from LEP2 require 
$m_{\tw_1}>103.5$~GeV and $m_{\tell_{L,R}}\agt 99$~GeV, 
resulting in some tension between slepton-mediated annihilation scenarios
and the WMAP/LEP2 data (see however Ref. \cite{king}
for some models where sfermion exchange remains as the dominant neutralino
annihilation channel in the early universe). 
As a result, the more generic prediction of constrained supersymmetric
models today in the LEP2 allowed parameter space is an {\it
overabundance} of CDM: see Eq.~(\ref{eq:susypred}). In fact, it is only
for special parameter choices where the neutralino annihilation cross
section is enhanced, or where co-annihilation with colored or charged
sparticles is important, that the model prediction is in accord with the
measured abundance in (\ref{eq:Oh2}).

The situation is exemplified in the extensively studied minimal
supergravity model (mSUGRA)\cite{msugra}. This model posits that the
minimal supersymmetric standard model (MSSM) is the correct effective
theory valid between mass scales $Q=M_{\rm GUT}$ to $Q=M_{\rm weak}$.
It is assumed that 
SUSY breaking in a hidden sector induces {\it universal} soft SUSY breaking
terms for visible sector fields via  gravitational interactions. The
effective Lagrangian for the visible sector, renormalized at a very high scale
$Q\sim M_{\rm GUT}$, is thus parametrized by
a common mass parameter $m_0$ for all
Higgs and matter scalars, a common mass $m_{1/2}$ 
for the $SU(3)$, $SU(2)$ and $U(1)$ gauginos,
and a common trilinear scalar coupling parameter $A_0$.  The gauge and Yukawa
couplings and soft SUSY breaking terms are then evolved via
renormalization group equations (RGEs) from $M_{\rm GUT}$ to $M_{\rm
weak}$, and electroweak symmetry is broken radiatively due to the large
top Yukawa coupling. The model is completely defined by the well-known
parameter set
\be
m_0,\ m_{1/2},\ A_0,\ \tan\beta\ \ {\rm and}\ \ sign(\mu ),
\ee
where $\tan\beta$ is the ratio of Higgs field vacuum expectation values
$\tan\beta =v_d/v_u$, and $\mu$ is the superpotential Higgs mass term,
whose magnitude (but not sign) is determined by the electroweak symmetry
breaking minimization conditions.

The region of low $m_0$ and low $m_{1/2}$ 
(the so-called bulk region) of the mSUGRA model
where neutralino annihilation via slepton 
exchange occurs\cite{bulk}, is nearly ruled out as already noted above. 
This leaves only several
surviving regions in accord with (\ref{eq:Oh2})\cite{bb}: (1)~co-annihilation
regions at low $m_0$ where $m_{\ttau_1}\simeq m_{\tz_1}$\cite{stau,isatools}, 
or at particular $A_0$ values where $m_{\tst_1}\simeq m_{\tz_1}$\cite{stop},
(2)~resonance annihilation regions such as the $A$-funnel\cite{Afunnel}, 
where $2m_{\tz_1}\sim m_A$, $m_H$, and the $A \ (H)$-resonance
enhances the neutralino annihilation rate, or the extremely narrow light 
Higgs corridor, where $2m_{\tz_1}\simeq m_h$,\cite{drees_h}
and (3)~the hyperbolic branch/focus point region (HB/FP) at large $m_0$, 
where $\mu$ becomes small and the neutralino acquires a significant 
higgsino component\cite{hb_fp}, which enhances its
annihilation rate into vector bosons.
Aside from these regions, most of the parameter space of the mSUGRA
model is ruled out because the sparticle mass scale is too high
resulting in a suppression of the annihilation rate and 
a corresponding over-abundance of CDM. Finally, we
note that in mSUGRA, the $\tz_1$ is dominantly bino-like over all of parameter
space, with the exception being the HB/FP region, 
where it picks up a significant higgsino component, and becomes
mixed higgsino dark matter (MHDM).

While the mSUGRA model serves as an economic paradigm for SUSY
phenomenology in gravity-mediated SUSY breaking models, the assumption
regarding universality at $Q=M_{GUT}$ is not well-motivated theoretically, and
models with non-universal soft terms should be considered. In fact,
patterns of non-universality generically arise in many SUSY GUT and
string model incarnations.  But what theoretical template is then
suitable? In this report, we will maintain the phenomenological successes of
supersymmetric models, while extending the parameter space to allow
various patterns of non-universality.  Motivated by the successes of gauge
coupling unification and the observed consistency
of the light Higgs mass prediction ($m_h\alt 135$~GeV) in the MSSM with precision electroweak measurements, we
maintain the assumption that the MSSM is the correct effective theory
between $M_{\rm GUT}$ and $M_{\rm weak}$.  We also preserve the beautiful
mechanism of radiative electroweak symmetry breaking (EWSB) triggered 
by the large top quark Yukawa coupling and associated
renormalization group (RG) running of soft parameters from a high scale
such as $M_{\rm GUT}$ to the weak scale $M_{\rm weak}$.  We assume
degeneracy of matter scalar soft terms equal to $m_0$ at $M_{GUT}$ in
order to suppress unwanted
flavor-changing neutral current (FCNC) processes; Higgs boson soft mass
parameters may, however, be assumed to be different from $m_0$.  Also,
even in grand unified models
the three gaugino masses need not be unified at $M_{GUT}$ (since SUSY
breaking vevs need not necessarily respect the GUT symmetry). 
Finally, we assume standard Big Bang cosmology with the lightest neutralino
as a thermal relic making up the bulk of CDM in the universe: {\it i.e.}
$\Omega_{\tz_1}h^2\sim 0.11$.

With the increased freedom in the GUT scale parameter space that is now
possible with the relaxation of the various universality assumptions, we
will be able to find scenarios such that the $\tz_1$ gains a {\it
partial} wino or higgsino component: {\it i.e.} just enough to fulfill
the CDM relic density measurement (\ref{eq:Oh2}).  Models of this sort
with mixed higgsino dark
matter\cite{ellis_nuhm,drees2,nuhm1,belanger,mamb,lm3dm} (MHDM) or mixed
wino dark matter\cite{bn,winodm} (MWDM) have been collectively dubbed
models of a ``well-tempered neutralino'' in Ref. \cite{adg}. 
While tempering will vary the neutralino composition to attain the 
measured relic density, it is alternatively possible to vary the
neutralino mass: in this case, agreement with (\ref{eq:Oh2})
may arise via resonant enhancement of the annihilation cross section,
via stop or stau co-annihilation, or via the recently suggested 
bino-wino co-annihilation (BWCA) mechanism \cite{bwca}.

Since we will be working only with model parameter choices that
completely saturate the WMAP measurement, we will be
naturally interested in the associated direct detection of dark matter
by underground experiments searching for relic neutralino-nucleus
collisions. The spin-independent neutralino-proton elastic cross section, as a
function of neutralino mass, serves as a figure of merit for direct
detection experiments, and experimental sensitivities are usually shown
as $\sigma (\tz_1 p)\ vs.\ m_{\tz_1}$, where effects of the specific
nuclear target have been extracted. This allows for a direct comparison
of the capabilities of detectors with different target
materials. Predictions of direct detection scattering rates have been
worked out by many groups for the mSUGRA model\cite{direct}, as well as
for models with non-universal soft
terms\cite{nezri,bottino,munoz}. Frequently, if a model point yields a
relic density {\it in excess} of the value ({\ref{eq:Oh2}), it also
tends to give extremely low direct detection rates, painting perhaps too
pessimistic a picture for direct detection searches, given the WMAP
constraint. By the same token, requiring a model with a well-tempered
neutralino which yields the measurement~({\ref{eq:Oh2}), the mechanism
which increases neutralino annihilation rates in the early universe may
also increase the direct detection rates.  This is  the
case if we start with a bino-like $\tz_1$, and then temper it by adding
just enough of either a higgsino component, and in some cases even a
wino
component, so as to saturate
Eq.~({\ref{eq:Oh2}).  Large direct detection rates are generally not
expected if the neutralino relic density is brought into agreement via
neutralino co-annihilation with other charged or colored sparticles, or even
via BWCA.

\section{Direct detection cross sections in models with a well-tempered 
neutralino }

We begin with a brief overview, in Fig. \ref{fig:dd}, of the current
experimental upper limits on the spin-independent neutralino-proton
cross section $\sigma_{\rm SI}(\tz_1 p)\ vs.\ m_{\tz_1}$, along with
projections for their upgrades and other proposed experiments.
Currently the most stringent upper limit on direct detection of
neutralinos has been obtained by the CDMS experiment\cite{cdms}, a
cryogenic solid-state apparatus in the Sudan mine using $Si$ and $Ge$
targets. This limit, shown by the solid contour labelled CDMS, extends
down to cross sections of $\sigma (\tz_1 p)\sim 3\times 10^{-7}$~pb for
$m_{\tz_1}\sim 100$~GeV, and to about $2\times 10^{-6}$~pb for TeV
neutralinos.  Since the experiments are based on the measurement of the
recoil energy of the nucleus which, of course, reduces with increasing
neutralino mass, the limits become weaker (and ultimately saturate) for
neutralinos that are much heavier than the nucleus.  The goal of
CDMS~II, as well as of Edelweiss II\cite{edelweiss} and Cresst
II\cite{cresst}, is to achieve optimal sensitivities of $\sim 10^{-8}$~pb 
by 2007-2008, as shown in the curve labelled CDMS II. In the
long-term, CDMS plans to deploy 7 supertowers in the Sudbury mine site
in a set-up labelled SuperCDMS, and aims to achieve a sensitivity as low
as $10^{-9}$~pb by around 2012\cite{supercdms}.  At the same time, a
variety of projects are planned to construct large noble gas dark matter
detectors, using xenon\cite{zeplin,xenon}, argon\cite{warp,clean} and/or
neon\cite{clean} targets. Such detectors are cost-efficient, and can be
envisaged to reach the ton scale in target material, and in addition may
have neutron veto capabilities.  Without making any representation of
the feasibility of such detectors, we also show the projected reach in
Fig.~\ref{fig:dd} of the Warm Argon Project (WARP), 1400~kg detector,
which aims for a sensitivity of $10^{-10}$~pb, as indicative of this class
of detectors.
\FIGURE[!t]{
\epsfig{file=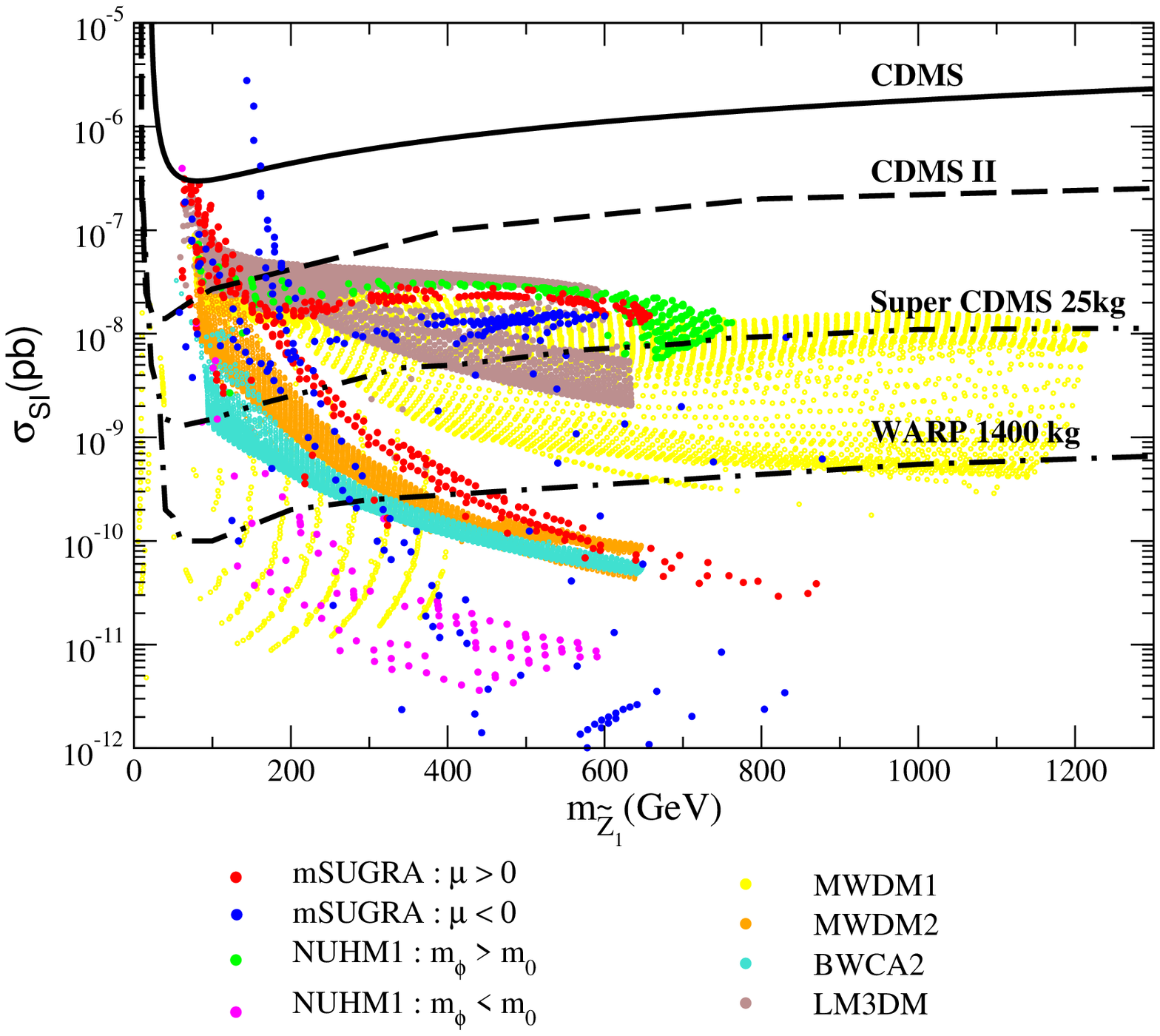,width=10cm,angle=-90} 
\caption{\label{fig:dd}
Spin-independent neutralino-proton scattering cross section versus
neutralino mass in various cases of models with a well-tempered neutralino.
We also show reach and projected reach of CDMS, CDMS II, SuperCDMS and
WARP 1400 kg detector. We take $m_t=171.4$ GeV.}}

In addition to the various experimental sensitivities in
Fig.~\ref{fig:dd}, we also show the expectation for the
neutralino-proton direct detection cross section in various models, but
only for parameter choices that give the neutralino relic density in
agreement with (\ref{eq:Oh2}). We take $m_t=171.4$~GeV in our analysis
\cite{tmass}.  Specifically, for each model we generate parameter space
points, and then use Isajet v7.74 for the calculation of the sparticle
mass spectrum and the associated neutralino relic
density\cite{isajet}. Then, for those points where the latter is
compatible with its observed value, we extract the spin-independent
neutralino-proton scattering rate from the IsaReS subroutine\cite{bbbo}
(a part of the Isatools package that includes the evaluation of relic
density \cite{isatools} and other low energy observables) and plot it in
Fig.~\ref{fig:dd}.
We begin by showing for reference 
the expectation for the direct detection cross section
for the paradigm
mSUGRA model for $\mu >0$ (red points) and $\mu <0$ (dark blue points).
We scan over $m_0:0-5$~TeV, $m_{1/2}:0.1-2$~TeV, $\tan\beta =10,\ 30,\
45,\ 50,\ 52$ and 55, and take $A_0=0$.
For both signs of $\mu$, we see that points form two distinct
branches. For positive $\mu$ (red points), the branch which extends 
from $\sigma\sim 4\times 10^{-7}$
pb at low $m_{\tz_1}$ to values below $10^{-10}$~pb for large
$m_{\tz_1}$ and is formed by bulk/stau co-annihilation and $A$-funnel
points.  The HB/FP points where $\tz_1$ forms MHDM lie on the
upper branch at roughly constant $\sigma\sim 2\times 10^{-8}$
pb\cite{bbbo}.  In this region, as higher values
of $m_0$ and $m_{1/2}$ are probed, an increasingly mixed bino/higgsino
$\tz_1$ is needed to maintain accord with (\ref{eq:Oh2})\cite{bkpu}.  As
a result, the neutralino has a slightly {\it increasing} direct
detection cross section. We note that it is precisely in this HB/FP
region, which falls entirely within the range of Super-CDMS experiment,
that the gluinos and squarks can be very heavy so that the direct
detection of supersymmetric particles at the CERN LHC
\cite{susylhc,bbbkt,bbko} is the most difficult, even with $b$-tagging
capabilities of the LHC detectors \cite{mizukoshi}.  
We also show the mSUGRA expectation for $\mu <0$ (blue points).
%
The direct detection cross sections in the HB/FP branch are only
slightly lower than for positive $\mu$, and should be within the projected
detection capability of Super-CDMS. A striking feature is that the
direct detection cross section for the stau-coannihilation/$A$-funnel
branch falls below even the WARP sensitivity for neutralino masses
bigger than just 300~GeV. The suppression of the cross section for
negative $\mu$ for large squark masses (where direct detection is
dominated by $h$ and $H$ exchanges between the nucleus and the
neutralino) was also noted in Ref. \cite{negmu}, and is the result of
several contributing factors: the neutralino coupling to $h$ is smaller
for negative $\mu$ because of a cancellation between the $H_u^0$ and $H_d^0$
contributions to the coupling, there is a negative interference between
the tree-level $h$ and $H$ diagrams,\footnote{
Although $m_H \gg m_h$, for large values of $\tan\beta$ the $H$-mediated
contributions to neutralino-nucleon scattering remain significant. This is
because $h \sim H_u^0$ when $\tan\beta$ is large so that its coupling to
the strange quark (which makes the dominant tree level contribution) is
suppressed by the Higgs mixing angle, whereas $H\sim H_d^0$ has an
essentially unsuppressed coupling to $s$-quarks.} 
but most importantly, diagrams where the
Higgs bosons couple via the gluon content of the proton through quark loops
interfere negatively (positively) with tree level diagrams where the
Higgs bosons couple to the quark content of the proton when $\mu <0$
($\mu >0$). 
%

Next we turn to models with a well-tempered neutralino. In order to
avoid extremely lengthy computer scans, for the most part we fix
$A_0=0$, $\tan\beta=10$ and take $\mu >0$.

\subsection {NUHM1 model: small ${\bf \mu}$ case}

The first model we investigate is the one extra parameter non-universal
Higgs model (NUHM1)\cite{nuhm1}.  These models are inspired by $SO(10)$
SUSY GUTs, wherein matter superfields belong to the 16-dimensional
spinor representation of $SO(10)$, while Higgs superfields belong to the
10-dimensional fundamental representation. To avoid unwanted FCNC
effects, we retain $m_0$ as the common matter scalar mass parameter
renormalized at $Q=M_{GUT}$, but now allow an independent SUSY breaking
mass squared parameter
$m_{\phi}^2$, which can take either sign, for both $H_u$ and $H_d$
fields. It has been shown in Ref. \cite{nuhm1} that for any 
mSUGRA parameter subset of the NUHM1 parameter set,
dialing $m_{\phi}^2\gg m_0^2$ leads to
a {\it diminution} of $\mu^2$, resulting in an
increased higgsino-content of $\tz_1$ which can then become MHDM.
We
scan over the mSUGRA $m_0\ vs.\ m_{1/2}$ plane for $m_0:\ 0-2$~TeV,
$m_{1/2}:\ 0-1.5$~TeV with $A_0=0$, $\tan\beta =10$ and $\mu >0$, while
tempering the $\tz_1$ at every point by dialing $m_{\phi}^2>> m_0^2$
until $\Omega_{\tz_1}h^2\simeq 0.11$ is attained.  The associated
spin-independent direct detection cross sections are then plotted as
green dots.  We have checked that other values of $A_0$ and
$\tan\beta$ give qualitatively similar results.
We find that the direct 
detection cross sections lie along a band at $\sigma\sim 1-3\times 10^{-8}$ pb,
and so this example of a model with a 
well-tempered bino/higgsino neutralino lies 
almost entirely within the reach of SuperCDMS.

\subsection {NUHM1 model: ${\bf A}$-funnel case}

Within the NUHM1 model framework, 
it is also possible to obtain agreement with (\ref{eq:Oh2}) by 
dialing $m_\phi^2\ne m_0^2$ to large negative values.
In this case, the $\tz_1$ remains nearly pure bino, but $m_A$ drops until
$m_A \sim 2m_{\tz_1}$, so that the neutralino annihilation in the early
universe is resonance-enhanced, even at low $\tan\beta$\cite{nuhm1}. 
Since a pure bino does not couple to $h$ or $H$, we
do not expect a significant direct detection cross section except
perhaps when sfermions are also light.  For this case- where $m_A$ rather
than the -ino content of $\tz_1$ is varied- the expectation is shown
by pink dots extending to rather low values
below $10^{-10}$~pb, which is below the projected reach of even the WARP
1400~kg detector.

\subsection{LM3DM model with MHDM:}

In this model, instead of using non-universal scalars, we adopt
non-universal gaugino masses. In particular, by taking $M_1=M_2\equiv
m_{1/2}$ at the GUT scale, but by dialing $|M_3({\rm GUT})|<<m_{1/2}$,
we reduce the gluino and squark masses. This reduction in sparticle
masses feeds into a reduction in the magnitude of the $\mu$ parameter
via the coupled RGEs and the EW minimization conditions\cite{lm3dm},
resulting in MHDM consistent with (\ref{eq:Oh2}) together with
low gluino and squark masses relative to charginos and neutralinos.
This is called the ``low $|M_3|$ dark matter model''
(LM3DM). For each point in the mSUGRA $m_0\ vs.\ m_{1/2}$ plane
(again for $A_0=0$, $\tan\beta =10$ and $\mu >0$) we reduce $M_3 >0$ until
({\ref{eq:Oh2}) is satisfied, and show the corresponding
direct detection cross section in
Fig. \ref{fig:dd} as tan points. We see in this model that
there exists a dense upper band of cross sections where $|\mu|$ is small
enough to be MHDM, where $\sigma (\tz_1 p)\sim 2-5\times 10^{-8}$~pb
which again falls within the projected reach of SuperCDMS. There is also
a lower band of tan points with lower direct detection cross sections
where compatibility with (\ref{eq:Oh2}) requires only a small reduction
in $|M_3|$ because annihilation via relatively light sleptons in the low
$m_0$ region of the parameter space also helps to yield the WMAP value of the
CDM density. Although the neutralino is
bino-like, $\sigma(\tz_1 p) \agt 10^{-9}$~pb (within the WARP
1400~kg reach) in this region, presumably because of both a slightly
enhanced higgsino content and a lighter squark mass (relative to
mSUGRA). We have checked that these results are qualitatively independent of
the sign of $M_3$. However, if we increase $\tan\beta$ to 30, while the
upper band remains essentially fixed, the lower one becomes more diffuse
and extends down to about $2\times 10^{-10}$~pb for the highest values
of $m_{\tz_1}$.

\subsection{Mixed wino dark matter model (MWDM)}

In the MWDM model\cite{winodm}, we may either take $M_1({\rm GUT})$ as a
free parameter with $M_2=M_3\equiv m_{1/2}$ 
and raise it until, {\it at the
weak scale} $M_1\sim M_2$ (MWDM1), or we can fix $M_1=M_3\equiv m_{1/2}$
and lower the GUT scale value of $M_2$ until again, {\it at the weak scale},
$M_1\sim M_2$ (MWDM2).  In both cases, the near equality of weak 
scale bino and wino masses
(with  $|\mu |$ remaining large) results in an LSP that is a mixed
bino-wino state with only a small higgsino admixture.  The
resulting mixed bino-wino $\tz_1$ has an increased annihilation rate
into $W^+W^-$ pairs (via chargino exchange).\footnote{If $M_1=M_2$ at
the weak scale, the LSP is a photino with electromagnetic couplings to
the $\tw^{\pm}W^{\mp}$ system, and the corresponding cross section 
for annihilation to $W$ bosons is governed by 
electromagnetic interactions.} Co-annihilation effects may also be
important. 
Within the MWDM framework, we can take any point in the $m_0\
vs.\ m_{1/2}$ plane of mSUGRA, and pull either $M_1$ up or $M_2$ down in 
value until we get a MWDM particle with $\Omega_{\tz_1}h^2\simeq 0.11$.  

The direct detection cross sections from the WMAP-consistent points for
the MWDM1 model (where $M_1$ is raised) are shown in Fig. \ref{fig:dd}
as yellow points. The striking feature is that the upper edge of the band of 
cross sections is detectable by Super-CDMS for all values of
$\tz_1$ masses.
The direct detection cross
section remains large in this band primarily because of the enhanced
coupling of the lightest neutralino to $h$. This is due in part to the fact 
that as $M_1$ increases (while $\mu$ and $M_2$ stay essentially fixed), 
there is not only increased bino-wino mixing, 
but also bino-wino-higgsino mixing.
Moreover, the relative sign of the bino and wino components of
$\tz_1$ is negative (it is positive for a pure photino state) so that the
contribution from the neutral higgs-higgsino-bino and
higgs-higgsino-wino {\it add constructively} in the $h\tz_1\tz_1$
coupling\cite{wss}.
%
%
There is also a group of points with intermediate and low $m_{\tz_1}$
that lie outside of the main yellow band. For those points, the 
$\tz_1$ remains bino-like, and the WMAP
value of relic density is achieved through various stau-coannihilation or
A-funnel mechanisms. For these points, 
the above-mentioned cross section enhancements are absent. 
and direct detection of the neutralino would only
be possible (if at all) at WARP 1400~kg. We have checked that if
we increase $\tan\beta$ to 30, the upper portion of the main yellow band
remains qualitatively unaltered (if anything, for very low $m_{\tz_1}$,
the cross section increases slightly), whereas the lower edge of this
band now spreads down to $\sigma(\tz_1 p) \sim 10^{-10}$~pb, so some
points with $m_{\tz_1} \agt 500$~GeV may not be within reach of the WARP
1400~kg detector. 
For these points, an increase of $M_1$ increases $m_{\tz_1}$ until
the $A$-funnel is reached before $\tz_1$ becomes MWDM, so $\tz_1$ 
remains bino-like.
If instead, we take $\mu < 0$, the {\it entire} main
yellow band shifts down and has $\sigma(\tz_1 p) \le 10^{-8}$~pb, so
that super-CDMS becomes insensitive for $m_{\tz_1}\agt 500$~GeV, and a
significant number of points closer to the lower edge of the yellow band
fall below the reach of even the WARP 1400~kg experiment. 
%
%

In the MWDM2 case where $M_2$ is lowered relative to fixed $M_1$ and
$M_3$, we see that the direct detection cross section (shown by orange
dots) falls off more rapidly with $m_{\tz_1}$ than in the MWDM1
case. This is because gluinos and squarks, and hence $|\mu|$, are
typically larger for a fixed LSP mass: as a result, the higgsino content
of $\tz_1$ is reduced (relative to the MWDM1 case). Furthermore, the
wino content, while increased relative to mSUGRA, remains significantly smaller
than in the MWDM1 case; the smaller wino component means 
that co-annihilations with $\tw_1$ and $\tz_2$ are crucial 
in getting the right relic density, so in fact this case resembles
BWCA dark matter, and the $\tz_1$ is only slightly tempered.
The small  wino component of $\tz_1$ results in a smaller coupling of the
neutral Higgs bosons to $\tz_1$ pairs compared to the MWDM1 case.
The direct detection cross section, which becomes roughly comparable to that
of the lower branch of the mSUGRA model, and may lie beyond even the
reach of WARP 1400~kg for $m_{\tz_1}\agt 350$~GeV.

\subsection{Bino-wino co-annihilation (BWCA) dark matter model}

The last method we study to get the observed value of the relic density is
to allow $SU(2)$ and $U(1)$ gaugino masses 
with {\it opposite signs}, but with their weak scale magnitudes nearly
equal: $|M_1({\rm weak})|\sim |M_2({\rm weak})|$ \cite{bwca}.  In this
case, there is essentially no mixing between the bino and wino states
and the $\tz_1$ remains nearly a pure bino DM particle.  However,
since $m_{\tw_1}\sim m_{\tz_2}\sim m_{\tz_1}$, bino-wino co-annihilation
can lower the LSP relic density to its measured value if $M_1$ or $M_2$ are
appropriately adjusted.  In the BWCA1 scenario, $M_1< 0$ is
adjusted for any fixed values of $M_2=M_3\equiv m_{1/2}$, while for the
BWCA2 scenario, it is $M_2 <0$ that is adjusted with $M_1=M_3\equiv
m_{1/2}$.  Because $\tz_1$ remains bino-like, we expect the
neutralino-nucleon scattering rate via Higgs boson exchange diagrams to be
small, so that $\sigma(\tz_1 p)$ will be relatively small unless squarks
are very light. This is borne out by our analysis.

In Fig.~\ref{fig:dd}, we show the direct detection cross section for the
BWCA2 model (light blue dots), where $-M_2$ is adjusted for a chosen
value of $M_1=M_3\equiv m_{1/2}$. In this case, 
we see that the scattering rates are
smaller than in the MWDM2 case, but may be observable at super-CDMS
(WARP 1400~kg detector) if the LSP is lighter than 160 (300)~GeV.
The smallness of this cross section is primarily
because the neutralino is essentially bino-like so that its higgsino
components, which are essential in order for it to couple to $h$ or $H$,
are very small. Squark mediated contributions are usually much
smaller. We have checked that the contributions from Higgs boson
couplings to quarks interferes constructively with the corresponding
(loop) contribution from its couplings to gluons, leading to the small,
but possibly observable cross section.

Although we do not show results for the BWCA1 model where $M_1 <0$ is
adjusted to give the relic density, we have checked that the range of
cross sections is qualitatively similar to the BWCA2 case except that
even for small values of neutralino mass, the direct cross section can
drop to well below $10^{-10}$~pb, so that it may not be detectable even
at the WARP 1400~kg experiment. This is essentially for the same reasons
(detailed above) that the cross section  can be small for
negative values of $\mu$ in the mSUGRA model. In the BWCA1 case, although
$\mu > 0$, the relative sign between the mass parameter $M_1$ of the
dominant gaugino component and $\mu$ is negative. The potential for the
destructive interference between contributing diagrams means that
although the cross section may be observable at the Super-CDMS (WARP
1400~kg) detector if $m_{\tz_1} \le 280$~GeV (400~GeV), direct detection
rates could be below the sensitivity of the WARP 1400~kg experiment even
for very low values of $m_{\tz_1}$.

\section{Conclusions}


In general scans over the LEP2-allowed portions of parameter space 
of supersymmetric models with GUT scale universality (such as mSUGRA), 
the predicted neutralino relic density is usually considerably above the WMAP 
measurement~(\ref{eq:Oh2}), while the direct detection rates are
pessimistically low. 
The predicted relic density matches its measured
value only if one is in a region of co-annihilation, 
of resonance annihilation, or of mixed higgsino dark matter annihilation
(such as the HB/FP region). 
In the last case, the predicted direct detection rates
for WMAP allowed points in parameter space are roughly constant with 
$m_{\tz_1}$ at a value $\sigma (\tz_1 p)\sim  10^{-8}$ pb.
Usually, as one proceeds to higher values of $m_{\tz_1}$, one has a
falling direct detection cross section. However, in the HB/FP region,
as $m_{\tz_1}$ increases, an increasingly larger higgsino component 
of $\tz_1$ is needed to maintain consistency with Eq.~({\ref{eq:Oh2}).
The large higgsino component 
also contributes to a direct detection cross section which is 
large and relatively stable against variations in $m_{\tz_1}$. 
The WMAP-allowed part of the HB/FP is 
an example of a region of parameter space where the composition of the
neutralino is tempered to give the observed value of the CDM relic
density. In the co-annihilation and resonance annihilation regions 
where {\it sparticle masses} are adjusted to 
give the measured relic density the neutralino remains a bino, and there
is no
enhancement of the direct detection cross section. 

In our study, we have examined a variety of models where we extend the
parameter space of the mSUGRA model, allowing either scalar mass or
gaugino mass non-universality, to obtain agreement with (\ref{eq:Oh2}). 
In well-tempered neutralino models where
the composition of the neutralino is dialed to give the
observed relic density, we typically get increased rates for 
neutralino direct detection because the coupling responsible 
for enhancement of the annihilation cross section
frequently also enhances neutralino-nucleon scattering. More to the
point, models of well-tempered neutralinos with mixed higgsino dark
matter yield neutralino-proton scattering cross sections that asymptote
to $\sim 10^{-8}$~pb for large neutralino masses, within the sensitivity
of the proposed 25~kg Super-CDMS upgrade of the CDMS experiment as
illustrated in Fig.~\ref{fig:dd}. 
Well-tempered neutralino models with mixed wino dark
matter may also yield detectable values of $\sigma (\tz_1 p)$ as illustrated by
the MWDM1 model in this figure. 
These asymptotic values of $\sigma(\tz_1 p)$ can serve
as target cross sections that proposed experiments should aim to
attain. In contrast, in the MWDM2 model, where 
the observed value of relic density is
obtained more via co-annihilation processes, the wino-higgsino content
of $\tz_1$, and hence the direct detection cross section, remains
relatively unenhanced.

In models where the mass -- and not the composition -- of the neutralino
is varied to give the observed CDM relic density via resonance
annihilation or via co-annihilation, the neutralino remains dominantly
bino-like.  In these cases, the direct detection cross sections do not
asymptote with increasing $m_{\tz_1}$, and we do not expect an
enhancement of the direct detection rate, except for small parameter
regions where sfermions are also very light.

In summary, we have shown that if relic dark matter consists
predominantly of stable neutralinos that have been thermally produced in
standard Big Bang cosmology, projections for the reach of 
direct dark matter detection experiments 
are substantially improved in supersymmetric models where
the composition of the neutralino is adjusted to give the observed relic
density. In this case, the neutralino will likely be detectable at
proposed experiments. Unfortunately, there is no analogous improvement
in the corresponding projections if the measured relic density is
obtained by adjusting sparticle masses instead of the neutralino
composition.

\acknowledgments

This research was supported in part by the U.S. Department of Energy and
by the National Science Foundation.

%

\end{document}